\documentclass{emulateapj}

\usepackage{amsmath}
\usepackage{graphicx}
\usepackage{subfigure}
\usepackage{multirow}

\begin{document}

\title{
Unusual Broad-Line MgII Emitters Among Luminous Galaxies in BOSS}

\author{
Benjamin Roig\altaffilmark{1},
Michael R. Blanton\altaffilmark{1},
Nicholas P. Ross\altaffilmark{2} 
}

\altaffiltext{1}{
  Center for Cosmology and Particle Physics, Department of Physics, New
  York University, 4 Washington Place, New
  York, NY 10003}
\altaffiltext{2}{
  Lawrence Berkeley National Laboratory, 1 Cyclotron Road, Berkeley, CA
  94270, USA}

\begin{abstract}
Many classes of active galactic nuclei (AGN) have been observed and
recorded since the discovery of Seyfert galaxies. In this paper, we
examine the sample of luminous galaxies in the Baryon
Oscillation Spectroscopic Survey (BOSS). We find a potentially new
observational class of AGN, one with strong and broad MgII
$2799$\AA\ line emission, but very weak emission in other normal
indicators of AGN activity, such as the broad line H$\alpha$,
H$\beta$, and the near-ultraviolet AGN continuum, leading to an 
extreme ratio of broad H$\alpha$/MgII flux relative to normal 
quasars. Meanwhile, these objects' narrow-line flux ratios reveal 
AGN narrow-line regions with levels of activity consistent with the 
MgII fluxes and in agreement with that of normal quasars. These AGN may
represent an extreme case of the Baldwin effect, with very low
continuum and high equivalent width relative to typical quasars, but
their ratio of broad MgII to broad Balmer emission remains very
unusual. They may also be representative of a class of AGN where 
the central engine is observed indirectly with scattered light.  These 
galaxies represent a small fraction of the total
population of luminous galaxies ($\simeq 0.1$\%), but are more likely 
(about 3.5 times) to have AGN-like nuclear line emission properties 
than other luminous galaxiess. Because MgII is usually inaccessible 
for the population of nearby galaxies, there may exist a related 
population of broad-line MgII emitters in the local universe which are 
currently classified as narrow-line emitters (Seyfert 2s) or LINERs.
\end{abstract}


\section{Introduction}

Astronomers have concluded that most, and possibly all, massive
galaxies contain supermassive black holes at their centers. The
accretion of gas onto these black holes emits light, in continuum and
line emission --- in such cases the black hole is referred to as an
active galactic nucleus (AGN). We believe the geometry of gas and dust
around AGN is complex, leading to substantially different
observational signatures depending on viewing angle and other factors.
In addition, the bolometric luminosity and spectrum depend strongly on
the amount of fuel available for accretion.

The Eddington ratio $R_{\mathrm{Edd}} \equiv
L_{\mathrm{bol}}/L_{\mathrm{Edd}}$ of an AGN is a useful
quantification of its activity.  For luminous quasars
$R_{\mathrm{Edd}} \sim 0.1$, but at low redshift most supermassive
black holes spend most of their time with much lower activity
($R_{\mathrm{Edd}} \sim 10^{-8}$--$10^{-3}$). These low-luminosity AGN
are thought to have a different physical structure and emission
mechanism. In particular, given estimates of the available mass for
accretion, it appears inevitable that they are radiatively
inefficient, which at least partly explains their low luminosity.

At low luminosities, several studies suggest that AGN do not exhibit
broad-line regions \citep{ho08a}. Although the effect in \citet{ho08a}
is probably dominated by Low Ionization Nuclear Emission-line Regions
(LINERs), whose status as AGN is controversial (\citealt{yan11a,
  sarzi10a}), the same effect is seen for Seyfert nuclei as well
(\citealt{elitzur09a}). In addition, low luminosity AGN lack the ``big
blue bump'' in the ultraviolet-optical range, thought to be indicative
of emission from the inner accretion disk, and are instead dominated
by mid-infrared energy output.  Finally, underluminous quasars exhibit
the Baldwin Effect in a number of broad emission lines: the equivalent
widths of these lines decrease with increasing luminosity
(\citealt{baldwin77a, boroson93a, thompson99a}).  This last trend
appears consistent with the low ultraviolet-optical emission for very
low luminosity AGN.

The Seyfert classification scheme, based on Balmer line widths,
differentiates galaxies according to the region emitting the observed
lines.  Broad lines, produced in the high velocity dispersion region
within a few parsecs of the black hole, are frequently obscured by a
presumed dust torus surrounding the AGN. Most of the time this 
dusty torus completely hides the broad line region --- however, sometimes 
the broad line region remains visible (at least partially) due to light 
scattering into our line of sight off of material near the AGN. However, 
in all 
cases, the narrow line region extends to 100s of parsecs, outside the 
dusty region of greatest obscuration. Therefore, observationally AGN 
fall into two rough classes, Seyfert 1 galaxies with both broad and 
narrow hydrogen lines, and Seyfert 2 galaxies with only narrow-line 
hydrogen emission.

Variations in the relative strength and visibility of the Balmer lines
have led some investigators to define more detailed subdivisions of
Seyferts. Seyfert 1.5 galaxies have moderate-strength broad H$\alpha$
and H$\beta$; Seyfert 1.8 have weak broad H$\alpha$ and H$\beta$; and
Seyfert 1.9 have weak broad H$\alpha$ and only narrow H$\beta$ (see
\citealt{osterbrock06a, ho08a}).

In this paper, we search the Baryon Oscillation Spectroscopic Survey
(BOSS; \citealt{dawson12a}) sample of target galaxies for
broad MgII 2799\AA\ emission. BOSS's large wavelength coverage and
its target redshift distribution permits coverage of the MgII 
line at its 2799\AA\ vacuum rest frame wavelength region for 
a large fraction of its objects.

We find a new spectral class of AGN with broad MgII
$\lambda$2799\AA\ emission, but apparently extremely low
near-ultraviolet continuum luminosity and broad-line Balmer emission.
These galaxies appear to be even closer spectroscopically to Seyfert 2
galaxies than the Seyfert 1.9s are --- with no visible broad H$\alpha$
or H$\beta$, but a broad component in MgII indicating AGN activity.

Section \ref{sample} describes our sample selection. Section
\ref{type} describes our comparison of these galaxies with other known
AGN.  Section \ref{summary} summarizes our conclusions about these
objects.

\section{Sample Selection and Line Fitting}
\label{sample}

BOSS (\citealt{dawson12a}) is part of the Sloan Digital Sky Survey III
(SDSS-III; \citealt{eisenstein11a}) and is designed to measure the
large-scale structure of the universe by observing many luminous
galaxies up to redshift $z \leq 0.8$.  In the process of doing
so, it takes spectra of a very large (1.5 million) sample of galaxies,
allowing the discovery of unusual objects. In addition, BOSS's
observed wavelength coverage (3600 to 10400 \AA) and redshift range
allows us to observe the MgII $\lambda 2799$ line (as measured in 
the vacuum) for a much larger sample than possible with the SDSS-II 
target galaxy sample (which had wavelength coverage between 3800 and 
9200 \AA\ and a redshift limit $z\leq 0.6$).

\begin{figure}[t!]
  \centering
  \includegraphics[scale=0.4]{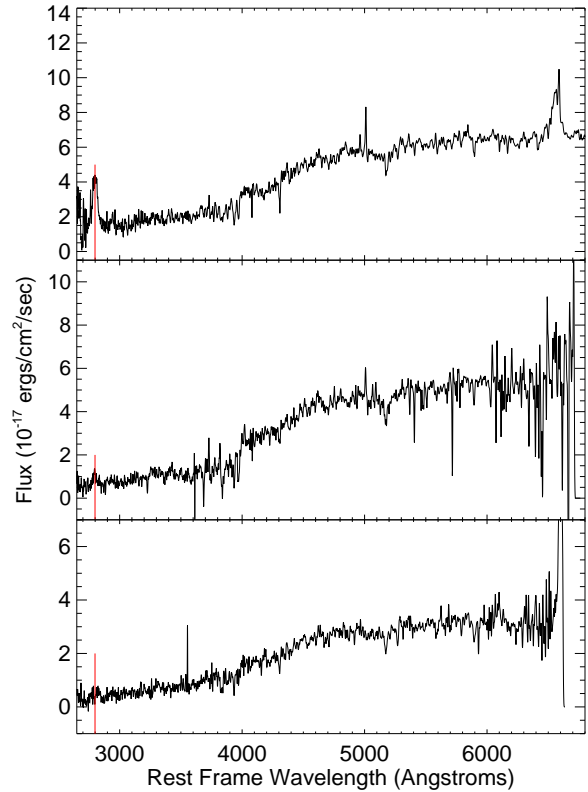}
  \caption{Three example BOSS target galaxies among the 293 broad-line 
    MgII emitters,
    showing of the variety of spectra accepted. The spectra are
    smoothed but otherwise unadjusted. The three spectra are ordered
    by MgII line strength, with the top panel the strongest and the
    bottom panel the weakest. The red line indicates the wavelength of
    the MgII line for reference. MgII line widths vary from
    approximately 2000 km/s to 3000 km/s for the given objects.}
  \label{examples}
\end{figure}

We selected objects from BOSS galaxy targets observed prior to January
2011, originally picked from the {\tt v5\_4\_14} reductions but updated
to now use the {\tt v5\_4\_45} spectroscopic reductions corresponding
to Data Release 9 (\citealt{ahn12a}).  We restrict the sample to
redshifts, $0.35 < z < 1.1$, for which both MgII and H$\beta$ are
within the BOSS spectral window (around 250,000 objects).  Most of the
LRGs in the BOSS data have $z < 0.8$, so extending the limit to $z =
1.1$ only adds a small number of objects.

We locate strong, broad-line MgII 2799\AA\ emitters in these spectra
by fitting a model consisting of a stellar continuum plus emission
lines. The stellar continuum is composed of eleven Chabrier IMF
stellar models of differing ages (13 Gyear to 0.1 Gyear), as detailed
in \cite{bruzual03a}. We find the best-fit linear combination of these
models to the observed continuum spectrum, with all non-stellar
emission line regions masked. Next, we fit the residuals around each
individual line region independently, with a single Gaussian model,
whose width, center, and flux are allowed to vary within reasonable
limits.  In the region around H$\alpha$, instead of independently
fitting the emission lines, we fit the [NII] doublet and the H$\alpha$
lines simultaneously with three Gaussians. In the fits described here,
we model each line with just a single Gaussian regardless of whether
we expect both broad/narrow components to exist.

After we perform this fit, we select galaxies based on their MgII
$\lambda2799$\AA\ properties in the {\tt v5\_4\_14} reductions. We select 
all galaxies that have a signal-to-noise ratio of at least $3.5$ and 
$\sigma > 12$\AA\ in the rest frame after de-redshifting the spectra 
(corresponding to a velocity dispersion of approximately $1300$ km/s at
 $z=0$).As an important note, in the update to {\tt v5\_4\_45}, the 
inverse variance recorded in the spectra was decreased by approximately 
a factor of 2 in all spectra, rendering the criteria used for this 
selection inaccurate for later versions --- to select a similar sample, 
we would need to reduce the signal-to-noise criteria by a factor of about 
1.4. Finally, if the software was unable to correctly fit the MgII 
line in a candidate luminous galaxy, we rejected the object (indicated by 
the line width returning its maximum value, $\sigma = 30.0$\AA\ ) This 
selection method ensures that only galaxies with real MgII detections 
remain in the sample --- some real and weak lines may be thrown out 
(rejecting all objects with $\sigma = 30.0$\AA\ removes about 150 objects, 
many of which may not have passed the visual inspection performed 
afterwards anyway), but the remaining sample we study will be as close 
to pure as possible. With these criteria, we pre-selected around 700 
galaxies out of the BOSS sample.

\begin{figure}[t!]
  \centering
  \includegraphics[scale=0.4]{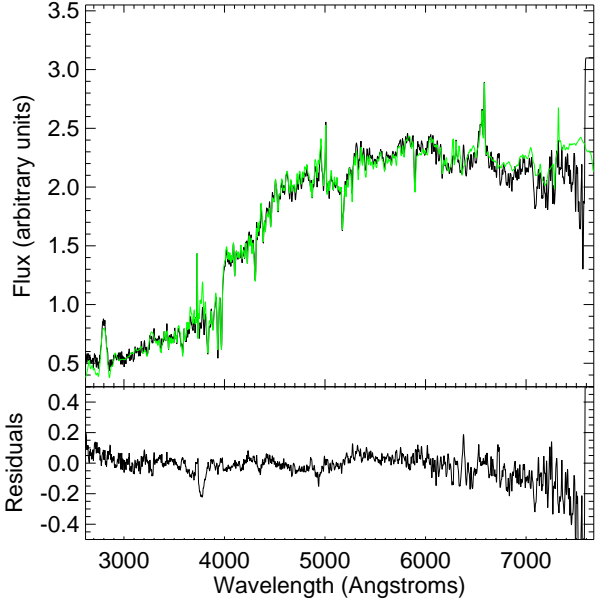}
  \caption{Top panel: Stack of the 162 luminous galaxiess in BOSS with 
    broad MgII detected and with wavelength coverage of both MgII
    $\lambda2799$\AA\ and H$\alpha$ ($0.35 < z < 0.57$). Our model fit
    to the stack is shown in green. Bottom panel: residuals of the
    stack minus the model.}
  \label{boss_stacks_ha}
\end{figure}

\begin{figure}[t!]
  \includegraphics[scale=0.4]{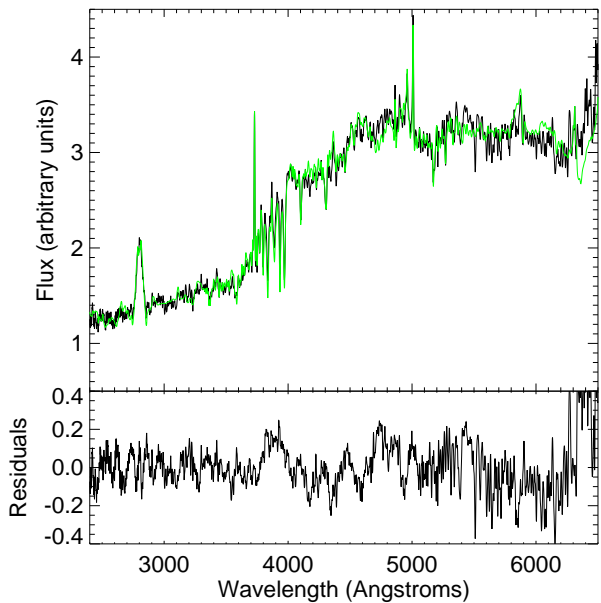}
  \caption{Similar to Fig. \ref{boss_stacks_ha} for the 131 luminous
    galaxies with broad MgII detected but no wavelength coverage for 
    the H$\alpha$ line, due to being too high redshift ($0.57 < z < 1.1$).}
  \label{boss_stacks_noha}
\end{figure}

Finally, we examined each of the spectra visually, inspecting the MgII 
line to see if it was a visually convincing strong line, and that
the fit to the data was reasonable. If the fit was unreasonable, or
the line extremely weak but broad (indicating a potential problem with
continuum subtraction), we dropped the galaxy from the sample. This
process eliminated about half of the previous galaxies, leaving us
with a sample of 293 galaxies that show broad MgII lines that were
both clearly visible and well-fit by our software. Some examples of
these are shown in Fig. \ref{examples} to show the varied types of
MgII lines that qualified as ``strong'' in this paper's sample. To
better compare properties discussed later in the paper, we then split
the sample in half based on redshift alone, into one set of luminous 
galaxies with $0.35 < z < 0.57$ (those with the H$\alpha$ line in the 
spectral coverage) and one set of luminous galaxiess with $0.57 < z 
< 1.1$ (those without the H$\alpha$ line).

Because these spectra are not individually high enough signal-to-noise
ratio to measure line fluxes and continuum levels precisely, we
de-redshifted and stacked the two subsets of the BOSS data. The
results of this stacking are shown (smoothed) as
Figs. \ref{boss_stacks_ha} and \ref{boss_stacks_noha} in black. We fit
the same model described earlier for the individual spectra to each
stacked spectrum; the figures show the result in green.  As the
residuals show, the fitting method is successful at matching the
spectra with a stellar continuum + Gaussian emission lines, with the
only major discrepancies arising on the red end of the spectra.

\begin{figure}[t!]
  \includegraphics[scale=0.4]{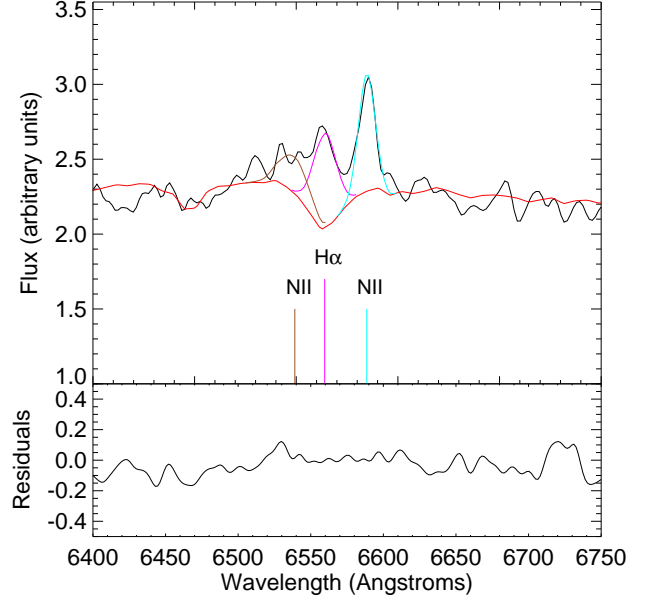}
  \caption{The fit from Fig. \ref{boss_stacks_ha}, zoomed in around
    the H$\alpha$ region. This plot also shows the individual
    gaussians used in fitting the two [NII] lines (brown and cyan) along
    with H$\alpha$ (magenta) and the base continuum flux (red). As
    this plot indicates, there is detected H$\alpha$ emission when the
    stellar absorption is accounted for. Although there is some 
    overlap between the lines, we can successfully fit the complex to
    obtain fluxes and widths for all three lines.}
  \label{boss_stacks_ha_hazoom}
\end{figure}

The spectra show very strong MgII, but the other common lines expected
in AGN are less obvious and only appear to contain narrow-line flux. For 
H$\beta$, the amount of measured flux
corresponds to what would be expected based on the H$\alpha/H\beta$
ratio. However, because of the underlying stellar absorption, the line
is only weakly apparent under visual inspection.  The H$\alpha$ and
[NII] complex is distinct in Fig. \ref{boss_stacks_ha} (or for a closer
view, see Fig. \ref{boss_stacks_ha_hazoom}), but only the narrow
components of these lines are detected (H$\alpha \simeq 250$ km/s,
H$\beta \simeq 150$ km/s). Finally, despite selecting strong MgII
emitters that are presumed luminous galaxies hiding AGN, the ultraviolet 
continuum is well described by stellar continuum, without much of a rise 
at the blue end as a typical quasar power law model would predict.

\section{What type of quasars are these?}
\label{type}

\subsection{Are the narrow lines consistent with AGN?}

\begin{figure}[t!]
  \centering
  \includegraphics[scale=0.42, trim=0 1.5in 0 1.5in, clip=true]{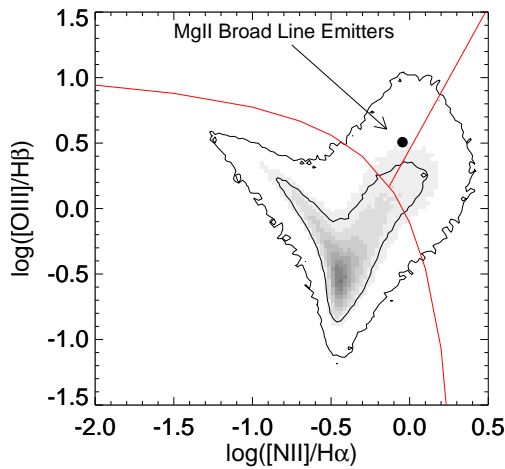}
  \caption{BPT diagram showing all SDSS DR8 galaxies (black
    background), overlaid with the stack of BOSS galaxies being
    considered (large black point). The uncertainties in the line
    ratios of the BOSS stack are approximately the size of the
    point. The red lines indicate the maximal star formation line
    dividing AGN from starforming galaxies from \cite{kewley01a} and
    the empirical line dividing LINERS and Seyferts from
    \cite{kauffmann03a}. Our stack lies in the predicted Seyfert
    section, although relatively close to the division between
    Seyferts and LINERs.}
  \label{bpt}
\end{figure}

With the sample of luminous galaxies selected from BOSS identified, 
the question now becomes ``what type of quasars are these?'' A first 
step in understanding
these objects is making a BPT diagram (as in \citealt{baldwin81a}) with
the subset that have spectra covering H$\alpha$. This is shown in
Fig. \ref{bpt}, along with the standard theoretical divisions from
\cite{kewley01a} and \cite{kauffmann03a}. The stack falls into a
Seyfert classification, though it lies rather close to the
conventional dividing line between Seyferts and LINERs. However, as 
Table \ref{lrg_classification} shows, many of the objects do not have 
at least one of the four lines measured, either due to low 
signal-to-noise in the individual spectra or redshift shifting the 
line out of the spectral range. For the objects with all lines, though,
it's clear that AGN-like emission (either Seyfert or LINER) is more
common in these MgII emitters than in the luminous galaxy population as 
a whole.

\begin{table}[t!]
  \caption{Classification of BOSS Target Galaxies in this Paper}
  \begin{tabular}{p{3cm}| c | c | c}
& All Luminous Galaxiess & MgII Emitters & \\ \hline
Total Number & 252228 & 293 \\
Missing 1+ BPT Line & 206849 (81.9\%) & 171 (58.4\%) \\
Starbursts & 26843 (10.6\%) & 44 (15.0\%) \\
LINERs & 13770 (5.5\%) & 58 (19.8\%) \\
Seyferts & 4826 (1.9\%) & 20 (6.8\%) \\
  \end{tabular}
  \tablecomments{The breakdown of the luminous galaxies in this sample, 
classified based on the [NII] / H$\alpha$ ratio and the [OIII] / 
H$\beta$ ratio. This table includes both the original set of BOSS 
target galaxies considered as well as the final selected sample to compare 
amounts of each type. It's interesting to note that while the average 
luminous galaxy from our BOSS sample is classified as a Seyfert, the most 
common object from those with all lines present is in fact a LINER.}
  \label{lrg_classification}
\end{table}

\subsection{Where are the broad Balmer lines?}

\begin{figure}[t!]
  \centering \includegraphics[scale=0.42, trim=0 2.2in 0
    0]{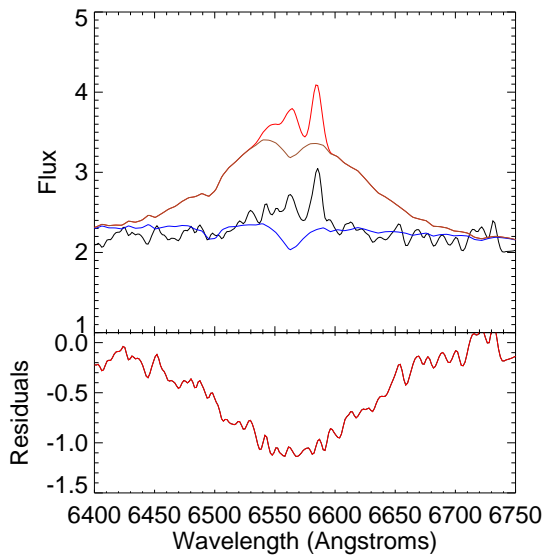}
  \caption{An example of what the H$\alpha$ line region would look
    like if our objects matched the broad H$\alpha$:MgII flux ratio 
    found in the Shen quasar sample, as shown in Table 
    \ref{shen_boss_lineratios2}, assuming the [NII] lines 
    are unchanged and the broad H$\alpha$ width is the same as our 
    measured MgII line width. The blue line indicates the continuum level. 
    The brown line shows the broad component and continuum, while the red 
    is the sum of all 4 Gaussians.}
  \label{broad_ha_added}
\end{figure}

Despite this evidence of the existence of the narrow-line region of a
quasar, the H$\alpha$ line and H$\beta$ line do not appear as strong
and broad as the Mg II line. In principle the broad component could be
present but not easily visible. In this section we will use line ratios
from SDSS quasars to predict what we expect for broad lines in our BOSS 
objects, and also compare various broad:narrow and broad:broad lines to 
search for significant discrepencies in the hope of further understanding 
our BOSS sample. First, to test whether these broad lines are present, we
considered the set of quasars in the DR7 catalog \citep{shen11a} with
both broad and narrow H$\alpha$ lines detected --- approximately 4600
objects. Assuming that the ratio of narrow H$\alpha$ flux to broad
H$\alpha$ flux is consistent in most quasars, and that the DR7 catalog
has objects comparable to the ones we are considering in BOSS, we can
predict what the broad H$\alpha$ flux would be in our BOSS stacked
spectrum from the observed narrow-line flux. We find that on average,
the SDSS quasars with both broad and narrow H$\alpha$ measured have a
narrow line to broad line flux ratio of at least 1:10 in H$\alpha$. When
we apply that same ratio to the observed H$\alpha$ narrow line in the
AGN in our sample of luminous galaxies in BOSS, we can quickly see that 
the broad line
emission (if present) would be clearly visible to us --- and so while
there may be small amounts of broad line flux present, it is certain
that these objects have a much smaller ratio of broad-to-narrow
H$\alpha$ than the Shen quasar sample. Fig. \ref{broad_ha_added} shows
what we would expect to see if we had the ``proper'' amount of broad
H$\alpha$ in these quasars given the measured narrow H$\alpha$. The key
to this is the standard [NII] doublet at 6550 and 6589 \AA, as the 
redder line in the doublet is clearly visible, forcing the blueward 
line to have a well-defined flux to be close to the normal ratio 
(approximately 3:1). However, the [NII] 6550 line is weak in our sample, 
and this constraint on [NII] 6550 strongly limits the strength of any 
broad feature underlying the H$\alpha$-[NII] complex.

\begin{figure}[t!]
  \centering
  \includegraphics[scale=0.42, trim=0 2.2in 0 0]{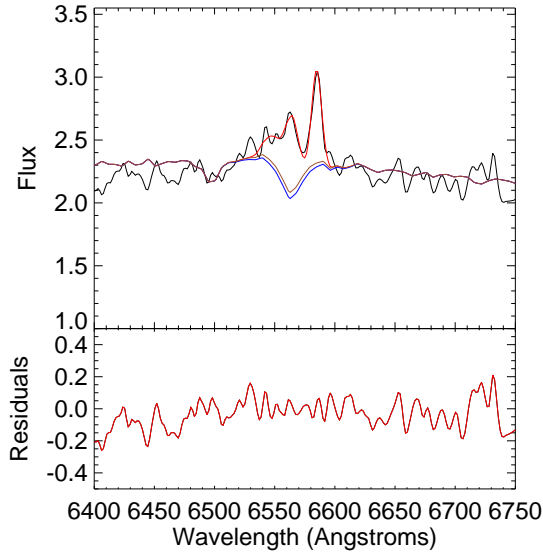}
  \caption{The best-fit result when we try to incorporate a broad
    H$\alpha$ component into our model, allowing the stellar population 
    portion to change as needed to maximize the amount of broad H$\alpha$ 
    that could be in the model. As is clear, very little broad
    H$\alpha$ is allowed in order to fit the spectrum well, despite
    what predictions from previously studied quasars would suggest we
    find. As in Fig. \ref{broad_ha_added}, the blue line shows the
    continuum, the brown the broad H$\alpha$ plus continuum, and the
    red the full model, and here the blue and brown lines are nearly
    identical.}
  \label{broad_ha_fit}
\end{figure}

\begin{table}[t!]
  \caption{Line Ratios of Quasars and BOSS Luminous Galaxies: 
    Narrow line:MgII flux ratios}
  \begin{tabular}{p{2.1cm} | c | c | c | c}
&\multicolumn{2}{|c|}{0.35 $<$ z $<$ 0.57} & z $>$ 0.57 \\
Objects & N. H$\alpha$/MgII & [OIII]/MgII & [OIII]/MgII \\ \hline
BOSS galaxiess & 0.306$\pm$0.01 & 0.280$\pm$0.007 & 0.244$\pm$0.006 \\
SDSS quasars & 0.458 & 0.149 & 0.091 \\
SDSS quasars & & & \\ (high MgII EW) & 0.160 & 0.113 & 0.112 \\
  \end{tabular}
  \tablecomments{These line ratios refer to spectra in
    Figs. \ref{shen_boss_1} and \ref{shen_boss_2}. The first column
    only lists the narrow H$\alpha$ flux, not including any
    potential broadline components. The luminous galaxies in BOSS have, 
    for a given 
    MgII amount, more narrow [OIII] flux than the set of SDSS quasars, 
    but do not follow the same pattern for narrow H$\alpha$. The subset
    of SDSS quasars in the last line is the set with large ($> 149$
    \AA) equivalent width in the MgII line (to be discussed in
    Section \ref{sec:cont_tell}).  }
  \label{shen_boss_lineratios}
\end{table}

\begin{table}[t!]
  \caption{Line Ratios of Quasars and BOSS Luminous Galaxies: 
    Broad H$\alpha$ flux ratios}
  \begin{tabular}{p{2.1cm} | c | c | c | c}
&\multicolumn{2}{|c|}{0.35 $<$ z $<$ 0.57} \\
Objects & Broad H$\alpha$/MgII & Broad H$\alpha$/Narrow H$\alpha$  \\ \hline
BOSS LRGs & $<$ 0.101 &  $<$ 0.330 \\
SDSS quasars & 2.85 & 36.6 \\
SDSS quasars & & \\ (high MgII EW) &  1.29 & 42.8 \\
  \end{tabular}
  \tablecomments{These line ratios refer to spectra in
    Figs. \ref{shen_boss_1} and \ref{shen_boss_2}. Here we see strong
    discrepencies between the Shen catalog and our sample of luminous 
    BOSS galaxies, 
    with significantly less broad H$\alpha$ flux than would be 
    anticipated, exacerbated since we can only accurately give an 
    upper limit on the H$\alpha$ broadline flux. The subset of SDSS 
    quasars in the last line is the set with large ($> 149$ \AA) 
    equivalent width in the MgII line (to be discussed in Section 
    \ref{sec:cont_tell}).   }
  \label{shen_boss_lineratios2}
\end{table}

In addition to this test, we consider how much broad H$\alpha$ we
could possibly allow, to quantitatively determine how close our
narrow-to-broad ratio is to that in the Shen quasar sample.  We
adjusted our fitting routine to fit 4 Gaussians around the H$\alpha$ +
[NII] complex --- one for each [NII] line, and a broad and narrow
H$\alpha$. Because our fit was very good to the narrow lines, we fix
the widths of the two [NII] Gaussians and the narrow H$\alpha$
line. Since we believe the MgII line accurately traces the broad
line region as well, we fixed the width of the broad H$\alpha$ line to
that value, because both lines should be produced in a similar
environment. In the end, then, our fitting program only can adjust the
relative amounts of flux in the four Gaussians in its attempt to model
the spectrum.  We find in Fig. \ref{broad_ha_fit} that very little
broad H$\alpha$ can be hidden under the continuum.

Table \ref{shen_boss_lineratios} shows a comparison of some of the 
line ratios between the luminous BOSS galaxies and the SDSS quasars. 
We compared various
narrow:broad line ratios in our BOSS objects, the full set of SDSS
quasars, and the subselected set of SDSS quasars described later in
Sec. \ref{sec:cont_tell}. Because we already found a difference in the
ratio of narrow H$\alpha$ to broad H$\alpha$, we examined whether the
discrepency persisted in other ratios. Table
\ref{shen_boss_lineratios} shows a far smaller, but by no means negligible,
level of discrepancy.  We consistently find a modestly higher flux of
[OIII] narrow-line emission relative to broad MgII (qualitatively
similar to the high narrow-to-broad H$\alpha$ ratio, but a much
smaller quantitative effect).  We also find that considering a subset of 
SDSS quasars with similarly large MgII EW gives us much smaller flux 
ratios, as the amount of MgII flux has increased without a change in the 
other line on average. Based on this table, our luminous BOSS galaxies seem 
similar to ordinary SDSS quasars in their properties, except for their 
apparently very low broad H$\alpha$ flux.

\begin{figure}[t!]
  \centering
  \includegraphics[scale=0.42, trim=0 4.8in 0 0, clip=true]{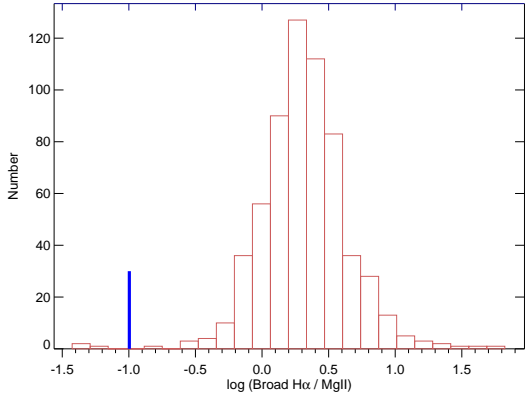}
  \caption{The distribution of the ratio of broad H$\alpha$ flux to MgII
  flux for the Shen catalog, all objects that contain both recorded line
  fluxes. The plot is presented as the log of this value. The luminous 
  BOSS galaxy stack we are comparing is shown with its value as a blue 
  line at log(broad H$\alpha$/MgII) $\simeq -1$, a distinct outlier.}
  \label{mgii_broadha}
\end{figure}

However, the more interesting results come in comparing broad:broad
line ratios, as Table \ref{shen_boss_lineratios2} does. Here we find
an extreme difference - the ratio of broad H$\alpha$ to MgII is
exceptionally small in our BOSS objects compared to the Shen sample, both
the selected subset and the entire sample. To show the magnitude of this,
Fig. \ref{mgii_broadha} shows the distribution of this ratio for all Shen
objects with both lines detected (the mean and standard deviation of
which are presented in Table \ref{shen_boss_lineratios2}). Only 7 or so
objects are as extreme in this flux ratio out of the nearly 650 objects in
the Shen catalog that have both lines detected, suggesting that this object
is far from the norm in this ratio. Similarly when comparing the broad
H$\alpha$ flux to the narrow H$\alpha$ flux, we find that the BOSS objects
are quite low in comparison to the Shen sample --- the ratio is about 
3:1 narrow to broad rather than the 1:30 narrow to broad found from the 
Shen quasars. These results combined, in addition to what we visually found 
by fitting the spectrum in Fig \ref{broad_ha_fit} suggests that the broad 
H$\alpha$ flux is strongly suppressed, an unexpected result given how 
strong and broad the MgII line is.

Most notable in the results from these two tables is lack of any dramatic 
discrepency between the ratios of broad MgII and narrow H$\alpha$ when 
compared to the SDSS quasars, combined with the large difference found between
broad H$\alpha$ and narrow H$\alpha$ discussed above. Together, these
results imply that the broad H$\alpha$ is very strongly suppressed
even with respect to broad MgII.

\begin{figure}[t!]
  \centering
  \includegraphics[scale=0.42]{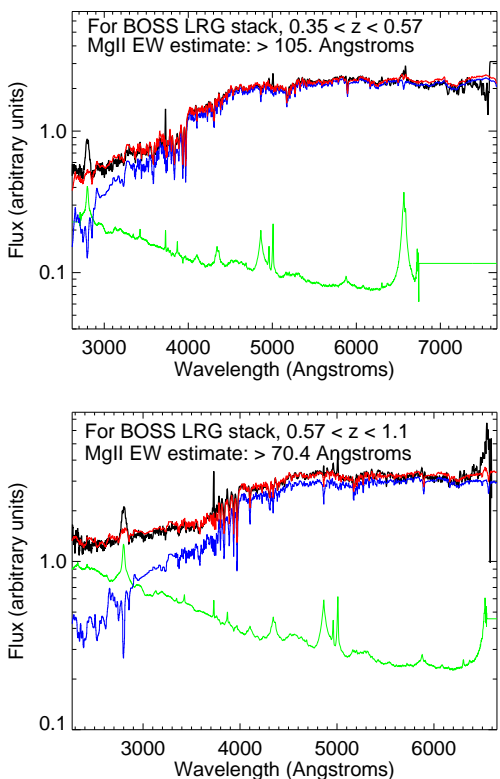}
  \caption{Model fit of a normal SDSS quasar and stellar continuum to
    luminous BOSS galaxies with broad MgII in both the $0.35 < z < 0.57$ 
    and $0.57 < z < 1.1$ redshift bins. The black line is the BOSS stack 
    we are
    fitting to, the blue line indicates the stellar continuum fit to
    that data, and the red line is the overall model of that continuum
    plus the SDSS quasar component. The green line below is the
    maximum amount of flux from the SDSS quasar component that the
    model allows at 95\% confidence. Based on the poor fit to the
    lines, these BOSS objects are extremely high MgII EW and not
    similar to the SDSS quasars.  }
  \label{shen_boss_1}
\end{figure}

\subsection{Where is the continuum flux?}

Next, we compared the two spectra in more detail to that of known SDSS
quasars, again taken from the catalog of \cite{shen11a}. To do this,
we created a ``typical'' SDSS quasar by splitting the quasars into the
same two groups based on spectral coverage of H$\alpha$, H$\beta$, and
MgII as described previously and then stacking them (for these
objects, the corresponding redshift ranges are $0.36 < z < 0.40$ and
$0.40 < z < 0.90$).  Because the BOSS target galaxiess were identified 
primarily as luminous galaxies and had a strong and easily fit stellar 
continuum component,
to best determine how much of the flux was from the quasar component
we took both the SDSS quasar stack and the stellar continuum models
from earlier (shown in Figs. \ref{boss_stacks_ha} and
\ref{boss_stacks_noha}), and combined them into one 12-component model
with which we refit our stacked spectra.  The model parameters reveal
how much continuum can be attributable to the stars in the luminous 
galaxy and how
much can be attributable to the quasar, as well as estimate or place a
limit on the equivalent width of MgII relative to the quasar
continuum.  Fig.  \ref{shen_boss_1} shows these results for the
H$\alpha$ and no-H$\alpha$ stacks.

The details of the quasar component of the best-fit model are not
particularly well constrained, and it is better to express the results
as an upper limit on the component that can be coming from a typical
quasar. To do so, we evaluated the 95\% confidence limit based on the
increase in $\chi^2$ as we increased the quasar amplitude. This limit
is shown as the green line in Figure \ref{shen_boss_1}.

\begin{figure}[t!]
  \centering
  \includegraphics[scale=0.42, trim=0 4.8in 0 0, clip=true]{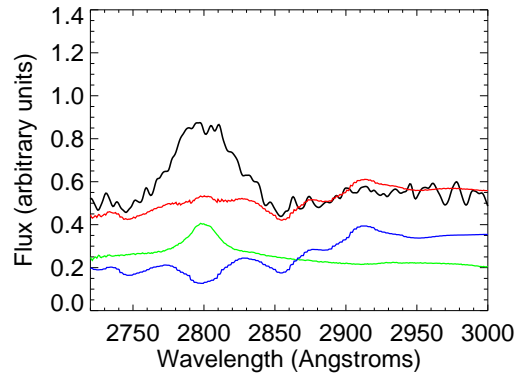}
  \caption{A zoomed-in view of panel 1 of Fig. \ref{shen_boss_1} to
    show how badly the MgII $\lambda2799$\AA\ line is fit by assuming
    the luminous BOSS galaxy spectrum to be composed of a stellar 
    component plus a
    standard SDSS quasar component.  Again the black line shows the
    BOSS stack, the blue line is the best-fit result of the stellar
    continuum, and the green line indicates the 95\% confidence
    maximum model of the quasar continuum. The red line represents the
    full model, the sum of the bright green and blue lines.  }
  \label{shen_boss_mgzoom}
\end{figure}

These figures suggest that these luminous galaxies, despite their broad 
MgII emission, do not have typical quasar spectra.  The ultraviolet
continuum in Figure \ref{shen_boss_mgzoom} is sufficiently
well-explained by the stellar emission such that the much bluer quasar
continuum is forced to low amplitude in our best-fit model (the
vertical scale is logarithmic in these figures).

Spectrophotometric calibration is a potential concern; the systematic
uncertainties in the ultraviolet due to calibration are likely to be
larger than the statistical errors.  To test for such issues issues,
we also attempted modeling the stacked spectra with the 12 components
described above, and two additional components, one constant and one
linear in wavelength, to allow the spectrum to tilt a bit. While this
did indeed produce a somewhat better fit with a greater contribution
from the quasar model, it also required that we have fluxing errors on
the order of a factor of two at the blue end of the spectrum (below 
4000 \AA\ mostly), far more than should be expected in these spectra (around 
10\% at most).

\begin{figure}[t!]
  \centering
  \includegraphics[scale=0.42, trim=0 4.8in 0 0, clip=true]{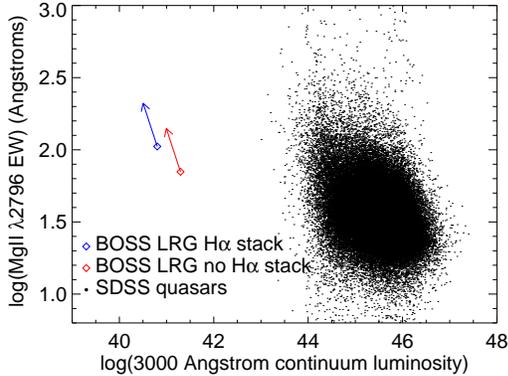}
  \caption{Given the 95\% confidence fits in Fig. \ref{shen_boss_1},
    this plot shows the limits on the MgII equivalent width in the
    luminous BOSS galaxies relative to the SDSS quasars. The SDSS 
    quasars exhibit the Baldwin Effect, a decrease in the equivalent 
    width as a function of quasar luminosity. As expected, the luminous 
    BOSS galaxy stacks are high EW, which may be related to their low 
    luminosity.  Indeed,
    they lie roughly along the predicted path of lower continuum
    luminosity and higher EW. The two points shown are the stacks in
    the redshift range $0.35 < z < 0.57$ (blue) and $0.57 < z < 1.1$
    (red).}
  \label{baldwineffect}
\end{figure}

\subsection{What do the continuum limits tell us?}
\label{sec:cont_tell}

Now noting that the AGN portion of the continuum is apparently very
low, we wanted to see if the large MgII flux is possibly an extension
of the Baldwin Effect as described in \cite{baldwin77a} in an extreme
case of very low continuum. We calculate the equivalent width (EW) of
the MgII $\lambda2799$\AA\ line by using the line flux found in our
line fitting procedure from earlier, and take the 95\% confidence
maximum quasar continuum as our continuum value. Fig. \ref{baldwineffect} 
shows that the luminous BOSS galaxies are indeed low continuum and high 
MgII flux.  
Because our measurement of the AGN continuum is an upper limit, we show 
the luminosity and EW in this plot as a 95\% confidence limit. Thus, 
although these luminous galaxies are certainly not similar spectroscopically 
to the Shen catalog of SDSS quasars, they do following a known trend for 
quasars and are conceivably just an extreme but not unknown case.

\begin{figure}[t!]
  \centering
  \includegraphics[scale=0.42]{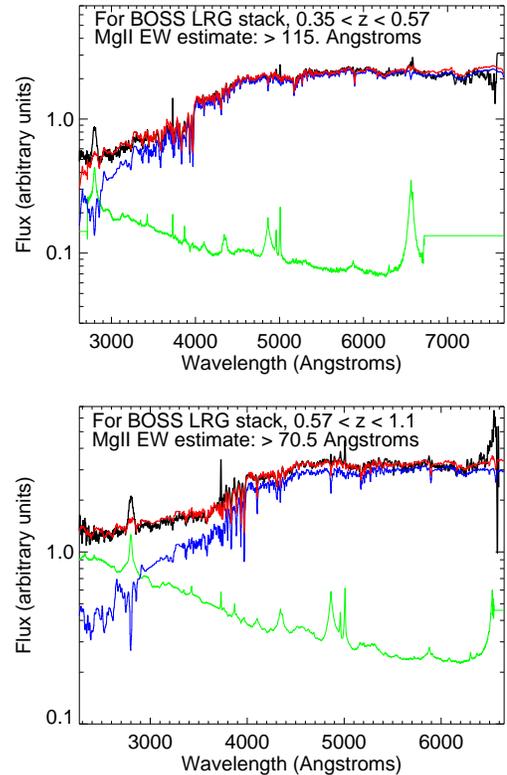}
  \caption{Similar to Fig. \ref{shen_boss_1}, but only stacking SDSS
    quasars with large ($> 149$ \AA) MgII EW. In this case, one of
    our two stacks has a better fit model (the bottom panel), while
    the other still remains a poor fit.}
  \label{shen_boss_2}
\end{figure}

To further examine this comparison of large MgII EW quasars to our BOSS 
galaxies, we then selected the subset of quasars from the Shen catalog
that have large MgII equivalent widths, potentially the closest
subset to the luminous BOSS galaxies we are examining. Few are as extreme 
as the luminous BOSS galaxies in both EW and luminosity, but if our objects 
are best
compared to quasars with large MgII EW, then performing the same fit
from before with only large MgII EW SDSS quasars should result in a
better match for both the continuum and emission lines. This
alternative fit is shown as Fig. \ref{shen_boss_2}. Here we see
progress in matching the spectra, especially in the bottom panel where
our model seems to explain the data relatively well, but note that the
top panel still shows a significant MgII flux deficit, suggesting
that possibly our objects require the most extreme MgII EW quasars to
be explained. Also of note is that this subset of quasars still does
not address the issues around the missing broad H$\alpha$. Other subsets
of the Shen catalog were considered (objects with small broad 
H$\alpha$/MgII ratio, objects with low Eddington ratio, and objects with 
the broadest MgII lines), but none of these approaches led to improved fit 
qualities in any of the lines or continuum regions.

\begin{figure}[t!]
  \centering
  \includegraphics[scale=0.42, trim=0 1.8in 0 2in, clip=true]{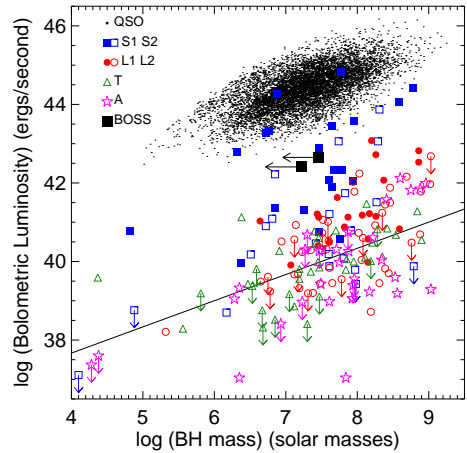}
  \caption{Comparison of the BOSS galaxies with broad MgII to the black hole
    masses and bolometric luminosity data presented in \cite{elitzur09a}.
    As labeled, the colors are the same as their original plots, with the 
    2 stacks of luminous BOSS galaxies overlayed as black
    points. The maximum BH mass is calculated given the maximum
    continuum fit in Fig. \ref{shen_boss_1} and represents an upper
    limit. The uncertainty in the bolometric luminosity is from the
    uncertainty in the [OIII] line fit, and is approximately the size
    of the data point.}
  \label{blr_disp_1}
\end{figure}

\begin{figure}[t!]
  \centering
   \includegraphics[scale=0.42, trim=0 1.8in 0 2in, clip=true]{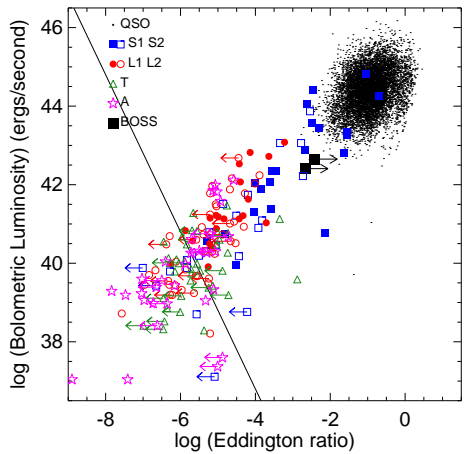}
  \caption{Comparison of the BOSS galaxies to the Eddington ratios and
    bolometric luminosities presented in \cite{elitzur09a}. As
    labeled, the are the same as their original plots, with the 2
    stacks of luminous BOSS galaxies overlayed as black points. The minimum
    Eddington ratio is calculated given the maximum continuum fit in
    Fig. \ref{shen_boss_1} and represents a lower limit. The
    uncertainty in the bolometric luminosity is from the uncertainty
    in the [OIII] line fit, and is approximately the size of the data
    point.}
  \label{blr_disp_2}
\end{figure}

\subsection{Should we expect a broad line region?}

The broad line regions in these luminous galaxies appear unusual, with a 
very low broad-line flux in the Balmer lines; in addition, the AGN have a
remarkably low continuum.  \cite{elitzur09a} found that given a BH
mass, there was a bolometric luminosity below which no broad-line
regions were found in the H$\alpha$ and H$\beta$ lines. The question
arises whether, given the result of \citet{elitzur09a} and the
observed narrow-line fluxes, we should have expected our MgII
emitting luminous galaxies to have broad-line regions, or whether
\citet{elitzur09a} would have predicted no broad line region.

To do so, we use the MgII line and $3000$\AA\ continuum luminosity to
estimate a black hole (BH) mass, following the prescription given in
\cite{wang09a}. Our $3000$\AA\ continuum luminosity is obtained from
the 95\% confidence maximum quasar continuum in the fits displayed in
Fig. \ref{shen_boss_1}. This procedure yields an upper limit on the BH
mass. Once the BH mass is obtained, we calculate its Eddington
luminosity. Finally, we derive a bolometric luminosity from the [OIII] 
$\lambda5007$\AA\ line using the relation derived in
\cite{heckman04a}, and then compute the Eddington ratio used in Figure
1 of \cite{elitzur09a}. Note that as these measurements were derived
from a 95\% confidence continuum, they correspond to a maximal BH mass
and a minimum Eddington ratio respectively, and are plotted as limits
rather than data points. Figs. \ref{blr_disp_1} and \ref{blr_disp_2}
reproduce Figure 1 from \cite{elitzur09a} with our data added as large
black points, with arrows showing the direction of the minimal/maximal
uncertainties explained above.

All these estimates are made on the basis of the MgII line because
the other, more commonly used lines are much weaker. Based on our 
findings above, these objects are unusual, and therefore the black
hole masses and Eddington ratios estimates may be incorrect.  In fact, 
we present some evidence for these black holes masses being somewhat 
underestimated below. Nevertheless, given the bolometric luminosities 
from [OIII], the black hole masses would have to be larger than any 
known black holes in the universe to violate the relation given in
\citet{elitzur09a}.

These results suggest that the broad-line regions in these luminous 
galaxies are safely consistent with the luminosity limits presented in
\cite{elitzur09a}. However, it is notable that these luminous galaxies
are missing
this BLR evidence in H$\alpha$ and H$\beta$, which for most local
samples are the only accessible lines. It is possible that some
galaxies in the local Universe currently classified as Seyfert 2s
actually house a detectable MgII region --- potentially even some of
those below the threshold determined in \citealt{elitzur09a}.

\subsection{Radio properties}

Here we investigate the radio properties of our sample galaxies, and
compare to typical broad-line radio galaxies (BLRGs).
\cite{rafter11a} investigated radio properties of confirmed broadline
AGN in SDSS in an attempt to correlate radio loudness with other
properties. To compare our sample with theirs, we searched the FIRST
catalog for objects in our sample with FIRST detections. Very few
objects appeared in both --- only 4 distinct objects of the 293 had
matches in FIRST within 0.5 arcseconds that were brighter than 3 mJy
integrated radio flux. Four other objects were found with very weak radio 
flux and are also included, but they are barely above the detection limit 
--- clearly not strong radio emitters. Despite several similar optical 
spectra, our sample does not appear to be similar to the BLRGs discussed in
\cite{rafter11a} on the whole.

\begin{figure}[t!]
  \centering
   \includegraphics[scale=0.42, trim=0 4.8in 0 0, clip=true]{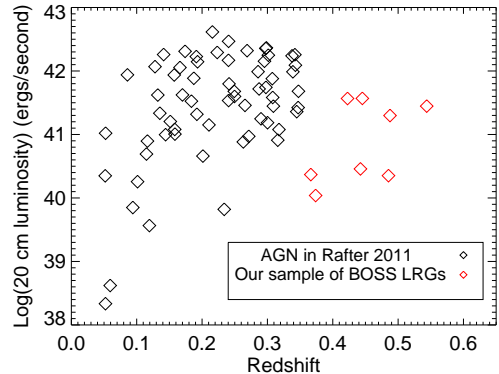}
  \caption{The redshift-radio luminosity plot showing our sample of
    luminous BOSS galaxies with detectable radio emission in FIRST 
    along with the
    sample discussed in \cite{rafter11a}. Our objects sit at higher
    redshift, indicating the possibility that the very limited
    detections were due to flux limits, but this plot suggests that
    this is not the case as we do not find the highest brightness
    radio sources anywhere in our 293 galaxies as in the Rafter
    sample, which we would expect if our galaxies were also
    BLRGs.}
  \label{radiocompare}
\end{figure}

However, it is worth noting that the objects that do have radio
detections are of similar luminosities as the sample in
\cite{rafter11a}. Fig. \ref{radiocompare} shows both sets of objects
plotting redshift versus luminosity. Since our objects are higher
redshift, it is possible that the reason fewer are detected was a
selection effect in FIRST; this appears to not be the case, however,
because we are missing objects in the high luminosity range as well.
Although our objects may end up being physically similar to typical
BLRGs, they are typically lower radio luminosity.

\subsection{Position relative to the $M_{\mathrm{BH}}$-$\sigma$ relation}

\begin{figure}[t!]
  \centering
  \includegraphics[scale=0.46]{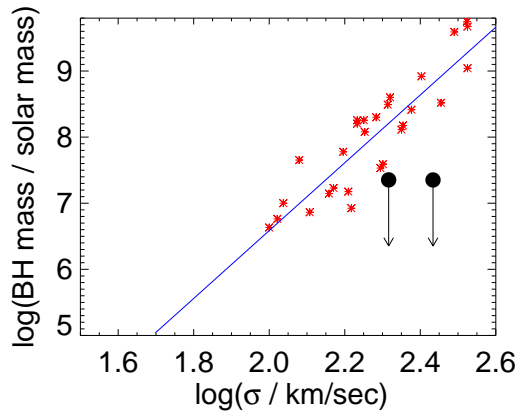}
  \caption{Our data relative to the $M_{\mathrm{BH}}$-$\sigma$
    relation, with our estimated upper limit for the black hole mass
    plotted as large black circles. Data from \cite{graham11a} is
    shown in red along with the best-fit line from the same paper in
    blue. Our objects are at best on the low end of the distribution,
    even given the intrinsic spread, and potentially much lower
    depending on the exact AGN continuum level. Either these black
    holes are extraordinarily low mass relative to their host
    galaxies, or existing relations between the MgII
    $\lambda2796$\AA\ line and black hole mass are flawed for objects
    like ours.}
  \label{mbhsigma}
\end{figure}

As a comparison to other black holes, we can place these stacked
objects from BOSS on the relationship between black hole mass and
stellar velocity dispersion. However, not every one of the luminous 
galaxies has a
recorded stellar velocity dispersion, so our data needed some slight
modification. To avoid biasing the estimated black hole mass due to a
systematic reason for an unmeasureable stellar velocity dispersion,
for this plot we added one more cut before stacking the objects,
measuring line and continuum fluxes, and calculating a black hole
mass. This new cut removed all objects that did not have an estimated
stellar velocity dispersion --- although this is a small percentage of
the total objects we kept, it has a non-negligible effect on the upper
limit BH mass estimate. Fig. \ref{mbhsigma} shows where our objects
lie, with the caveat as previously mentioned that we can only estimate
an upper limit on the black hole masses due to the very weak AGN
continuum.

Fig. \ref{mbhsigma} suggests quite strongly that for these luminous 
BOSS galaxies, the standard prescriptions (such as that of 
\citealt{wang09a}) for deriving a black hole mass from MgII do not 
work well. However, it is important to note that these values 
assume we are seeing the entire flux from the relevant lines. In 
section \ref{dustscatter}, we will note that perhaps our objects have 
scattered broad line flux and are not seen directly, and perhaps these 
estimates are too low for that reason. By not seeing the broad line 
region directly, we may not see the entirety of the MgII flux that is 
produced, and so our estimates depending on the fluxes in any broad 
lines could be underestimates. For normal (viewed directly) 
objects, MgII provides consistent (with H$\alpha$) BH masses (see 
\citealt{matsuoka13a, onken08a, mcgill08a}).  However, in our sample
we find much smaller black hole masses than the $M$-$\sigma$ relation
predicts. If the MgII line remains accurate as a BH mass indicator in
these objects, they fall below the typical masses given the velocity
dispersions of the galaxies, by much more than the scatter in the
relationship.

\subsection{Dust properties in the narrow line region}

\begin{figure}[t!]
  \includegraphics[scale=0.42, trim=0 4.8in 0 0, clip=true]{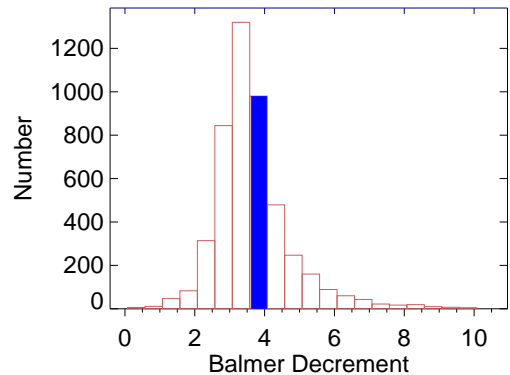}
  \caption{The distribution of Balmer decrements for all SDSS quasars
    from the Shen catalog that have both H$\alpha$ and H$\beta$
    detected.  (approx. 4800 quasars). The majority are around a value
    of 3--4, and the Balmer decrement for the BOSS stack is indicated
    as at the location of the blue bin.  The luminous BOSS galaxy stack's 
    Balmer decrement does not differ from the standard Balmer decrement 
    seen in many quasars.}
  \label{balmerdec}
\end{figure}

For the subset of the luminous galaxies in this sample observed between 
redshifts
$0.35$ and $0.57$, we can calculate the Balmer decrement in the
narrow-line fluxes, to estimate the degree of interstellar reddening
of the narrow-line region. A very reddened narrow-line region would be
suggestive of an unusual geometry or very dusty system, which might
allow us to attribute the lack of broad Balmer lines to these
properties of the system as well (though it would still be unusual
that MgII was not even more extincted).  Typical Seyfert galaxies
have a Balmer decrement H$\alpha$/H$\beta \sim 3$--$4$
(\citealt{osterbrock06a}) and for our objects we find a Balmer
decrements of $3.86 \pm 0.21$.  Figure \ref{balmerdec} shows the
distribution of Balmer decrements for typical SDSS quasars; the BOSS
objects are near the median of from this distribution. Of course, the
fact that the narrow-line region is not unusually affected by dust
does not rule out that the broad-line region still might be.

\subsection{Scattering and polarization in the broad line region}
\label{dustscatter}

In the previous section, we noted that no dust reddening was apparent in 
the narrow-line region. However, as we do not have any visible broad line 
H$\alpha$ or H$\beta$, we cannot make the same conclusion for the broad 
line region. It is possible that these objects have a geometry where there 
is a cloud (internal to the narrow line region, but external to the 
broad line region) obscuring the broad line region from our direct line of 
sight.Then, the broad lines we observe could be scattered instead of in 
direct line of sight of us, while the narrow line emission is unchanged.
Objects like this have been observed in the past --- see 
\cite{schmidt02a, schmidt07a, diseregoalighieri96a} among many others 
recording spectra of these rare but known objects.

Many quasars hidden in galaxies exhibit flat quasar spectra (see especially 
\cite{diseregoalighieri96a}) with prominent narrow lines. This is in contrast
to a directly viewed, very luminous quasar, which have a power law shape to
their continua, but similar to our observed objects (see Fig. \ref{shen_boss_1}). 
At the least, we cannot rule out a flat quasar spectrum to the same degree we can 
a standard luminous quasar spectrum --- objects such as those presented in 
the above papers would be better fits to our spectra. Unfortunately, none 
of these previously studied objects have the spectral coverage for both 
MgII and H$\alpha$, so we are unable to compare the broad line fluxes in 
both simultaneously. While none of them have such a strong galaxy component 
in their spectra, the quasar portions of the spectra are potentially very 
similar.

This flat spectrum can actually be hiding a more standard quasar spectrum
in polarized light, as seen in IRAS 09104+4109 (discussed in several papers: 
see \cite{hines93a} and \cite{hines99a} for details). This object's unpolarized 
continuum is nearly flat between 2500 and 6500 \AA, but when the polarized flux 
is examined the object reveals a clear quasar continuum with standard power law 
slope. It's quite possible that our objects resemble this, although we cannot 
confirm that without spectropolarimetry. The most important characteristic of
IRAS 09104+4019 is that it, like our objects, had no broad line emission recorded
in the Balmer lines in the initial detections in \cite{kleinmann88a} --- as the 
MgII line was not observed due to wavelength coverage --- until it was re-observed 
with broader band spectrophotometry by Hines and Wills. These followup observations
showed similar broad MgII emission to our objects. As IRAS 09104+4109 was 
eventually found to have broad Balmer line emission in a polarized spectrum, 
further investigation of our objects with spectropolarimetry may be useful to see
if similarities persist.

If the MgII line is being seen via scattered light, and our quasars remain
normal in terms of H$\alpha$ emission line production, then the efficiency 
of the scattering of H$\alpha$ must be very low relative to that of MgII. 
We see a large MgII flux already and no definitive broad H$\alpha$ --- so 
the scattering properties must be such to emphasize this discrepency (perhaps
similar to those in IRAS 09104+4019).

It would be very valuable to examine some of the strongest examples of 
these objects using spectropolarimetry to see if the MgII line is polarized 
as this hypothesis would predict, and to confirm the narrow lines are
unpolarized as well, as observed in IRAS 09104+4109.

\section{Summary}
\label{summary}

We have found a set of luminous galaxies in BOSS that exhibit unusual 
properties related to the MgII $\lambda2799$\AA\ line. Specifically we 
reach the following conclusions:

\begin{enumerate}
  \item This sample of galaxies has clear broad MgII line emission,
    but a much smaller broad H$\alpha$ flux ratio relative to MgII
    than typical for luminous quasars (by a factor of a few tens).
  \item This sample of galaxies have narrow Balmer line fluxes and [OIII] 
    flux ratios relative to MgII that are roughly typical for
    luminous quasars (within a factor of two or three).
  \item Galaxies with this property are an extremely small subset of
    the BOSS sample ($<0.1$\%).
  \item The continuum flux from the AGN in these luminous galaxies is 
    undetectably low, which means they must have a large equivalent width
    in the MgII line (possibly consistent with the Baldwin Effect seen for 
    more luminous quasars).
  \item Given the $M_{\mathrm{BH}}$-$\sigma$ relation, either the MgII 
    based black hole mass indicators used for more luminous quasars
    do not work for these lower luminosity objects, the black holes
    in these galaxies are small with respect to the expected mass
    given their host galaxy, or the light from the BLR is scattered
    such that we only see a fraction of the MgII emission.
  \item The galaxies appear to be low Eddington ratio ($<10^{-2}$)
    given the expected black hole mass.
\end{enumerate}

The physical nature of these objects is unclear.  Their lack of low
continuum flux may be related to whatever physical cause lies behind
the Baldwin Effect, which is itself still under debate.  Given how
uncommon these galaxies are, some unusual configuration around the
black hole may allow obscuration of the continuum but not the
broad-line region.  Many ideas have been proposed to explain how these
and other causes could contribute to the Baldwin effect (see
\citealt{green01a} for a discussion), such as disk inclination,
luminosity-dependent spectral energy distributions, or changing
optical thickness of clouds around the black hole, and it is possible
that one or more of these may be at play in our sample.

Why the broad-line Balmer emission remains so weak is also
unknown. Differences of this objects from luminous quasars in
metallicity or relative ionization fractions are unlikely to cause as
large a difference in flux ratios as we observe. We can only offer
up the wavelength-dependent scattering of the broad-line region as a possible
explanation as to why the broad Balmer lines could be missing and
the broad MgII line so strong.

Among known Seyfert 2 galaxies of lower stellar mass than these 
luminous galaxies, very few have had their rest-frame MgII $2799$ \AA\ 
region observed, because the vast majority of known galaxies are at 
low redshift and MgII 
is too blue to be observed. The population of broad-line galaxies
found here may exist in lower luminosity galaxies as well, in lesser
or greater abundance than found for the luminous BOSS galaxies.  Therefore, 
some fraction of galaxies currently classified as Seyfert 2s may have
broad-line regions that are observable, but only in the
as-yet-unobserved MgII line.

\acknowledgements

We thank Renbin Yan and Mike Eracleous for useful conversations during
the preparation of this paper. This work was partially supported by
NSF-AST 12211644 and NSF-AST-0908354.

We would also like to thank the referee for his very helpful
comments, especially directing us towards the class of AGN observed
via scattered light.

Funding for SDSS-III has been provided by the Alfred P. Sloan
Foundation, the Participating Institutions, the National Science
Foundation, and the U.S. Department of Energy Office of Science. The
SDSS-III web site is http://www.sdss3.org/.

SDSS-III is managed by the Astrophysical Research Consortium for the
Participating Institutions of the SDSS-III Collaboration including the
University of Arizona, the Brazilian Participation Group, Brookhaven
National Laboratory, University of Cambridge, Carnegie Mellon
University, University of Florida, the French Participation Group, the
German Participation Group, Harvard University, the Instituto de
Astrofisica de Canarias, the Michigan State/Notre Dame/JINA
Participation Group, Johns Hopkins University, Lawrence Berkeley
National Laboratory, Max Planck Institute for Astrophysics, Max Planck
Institute for Extraterrestrial Physics, New Mexico State University,
New York University, Ohio State University, Pennsylvania State
University, University of Portsmouth, Princeton University, the
Spanish Participation Group, University of Tokyo, University of Utah,
Vanderbilt University, University of Virginia, University of
Washington, and Yale University.

\bibliographystyle{apj} 
\bibliography{MgII_Emitters_BOSS}

\end{document}